\documentclass[aps,prl,twocolumn,citeautoscript,superscriptaddress,nopacs]{revtex4-1}

\usepackage{times}
\usepackage{amsmath,amssymb,amsfonts,mathrsfs,bm,feynmf,setspace}
\usepackage{dcolumn}
\usepackage{graphicx}
\usepackage{latexsym}
\usepackage{multirow}
\usepackage{color}

\begin{document}

\title{Harnessing the giant out-of-plane Rashba effect and the nanoscale persistent spin helix
       via ferroelectricity in SnTe thin films}

\author{Hosik \surname{Lee}}
\email[These authors contributed equally to this work.]{}

\affiliation{School of Mechanical, Aerospace and Nuclear Engineering, Ulsan National Institute of Science and Technology
(UNIST), Ulsan 44919, Korea}

\author{Jino \surname{Im}}
\email[These authors contributed equally to this work.]{}

\affiliation{Center for Molecular Modeling and Simulation, Korea Research Institute of Chemical Technology, Daejeon 34114, Korea}

\author{Hosub \surname{Jin}}
\email[Correspondence should be addressed to ]{hsjin@unist.ac.kr}

\affiliation{Department of Physics, Ulsan National Institute of Science and Technology (UNIST), Ulsan 44919, Korea}


\begin{abstract}
A non-vanishing electric field inside a non-centrosymmetric bulk crystal transforms into a momentum-dependent
magnetic field, namely, a spin-orbit field (SOF). SOFs are of great use in spintronics because they enable
spin manipulation via the electric field. At the same time, however, spintronic applications are severely
limited by the SOF, as electrons traversing the SOF easily lose their spin information. Here, we propose that
in-plane ferroelectricity in (001)-oriented SnTe thin films harness the Janus-faced SOF in a reconcilable way
to enable electrical spin controllability and suppress spin dephasing. The in-plane ferroelectricity produces
a unidirectional \emph{out-of-plane Rashba} SOF that can host a long-lived helical spin mode known as a persistent
spin helix (PSH). Through direct coupling between the inversion asymmetry and the SOF, the ferroelectric switching
reverses the out-of-plane Rashba SOF, giving rise to a maximally field-tunable PSH. Furthermore, the giant
out-of-plane Rashba SOF seen in the SnTe thin films is linked to the nano-sized PSH, potentially reducing spintronic
device sizes to the nanoscale. We combine the two ferroelectric-coupled degrees of freedom, longitudinal charge
and transverse PSH, to design intersectional electro-spintronic transistors governed by non-volatile ferroelectric
switching within nanoscale lateral and atomic-thick vertical dimensions.
\end{abstract}

\maketitle

The spin-orbit field (SOF) induced by inversion asymmetry has been an important factor in
spintronics~\cite{Zutic2004,Kato2004,Nitta1997} ever since the development of the Datta--Das spin field-effect
transistor~\cite{Datta1990,Koo2009,Chuang2015}, which harnesses the coherent spin precession controlled
by the Rashba SOF~\cite{Bychkov1984}. Although not unheard of, however, the SOF is rarely compatible with
long spin lifetimes because it breaks the spin rotational symmetry and suffers from fast spin decoherence
in a diffusive transport regime~\cite{Dyakonov1972}. Thus, acquiring electrical spin controllability while
preserving the long spin coherence has been elusive. As an exceptional example, the unidirectional SOF emerges
from the interplay between Rashba and Dresselhaus SOFs in III-V semiconductor quantum wells of general crystal
orientation and enables coherent spin manipulation by hosting the persistent spin helix (PSH), with a lifetime
of a few nanoseconds~\cite{Schliemann2003,Bernevig2006,Koralek2009,Walser2012,Schliemann2017,Kammermeier2016}.
However, the small and finely tuned SOF strengths in III-V semiconductor heterostructures have presented
a bottleneck in practical applications. Satisfying the stringent conditions for fine-tuning the Rashba and
Dresselhaus SOFs can complicate the attainment of the unidirectional SOF and the corresponding PSH. Even after
achieving the unidirectional SOF and PSH, efficient spin manipulation via the electric field is limited by the
small values and narrow ranges available to the Rashba and Dresselhaus terms. The small SOF then generates
a micrometer-long pitch in the PSH, hampering the ability to fabricate small spin transistors for high-density
and scalable spintronic devices~\cite{Sugahara2010}.

\begin{figure}
 \centering\includegraphics[width=8.5cm]{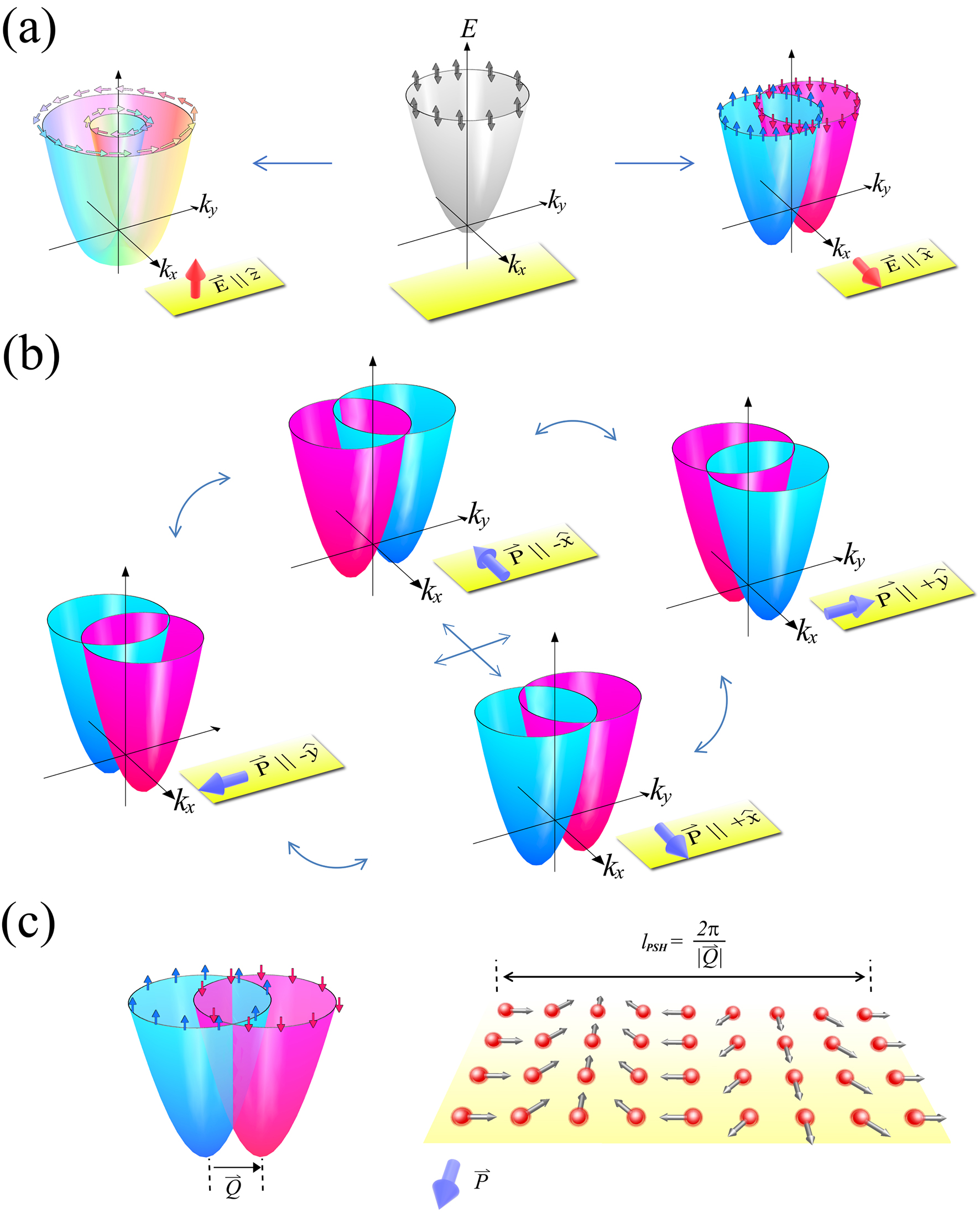}
 \caption{(a) In a two-dimensional electron system, the conventional and out-of-plane Rashba spin-orbit fields (SOFs)
 appear under the out-of-plane and in-plane electric fields, respectively. (b) The out-of-plane Rashba band induced by
 the in-plane ferroelectricity represents the maximally controllable SOF, depending on the ferroelectric configurations.
 (c) The out-of-plane Rashba SOF shifts the spin-up and spin-down parabolic bands by a constant vector $\vec{Q}$, which
 hosts the long-lived PSH with a spatial periodicity $\frac{2\pi}{|\vec{Q}|}$.}
 \label{fig:1}
\end{figure}

In pursuit of a better spin transport device, the present work sought an easily accessible and controllable PSH
with rapid precession in space and a slow decay in the time domain. The ideal target system should exhibit
a unidirectional, field-tunable, large SOF driven by a single inversion asymmetry. Instead of dealing with the
fine-tuned Rashba and Dresselhaus terms, the in-plane electric field in a two-dimensional electron system can
simply produce a unidirectional SOF. Given that the SOF generated by the constant electric field is expressed by
\begin{equation}\label{Rashba}
\vec{\Omega}_{\textrm{SOF}}(\vec{k})=\alpha(\hat{E}\times \vec{k})
\end{equation}
up to the leading order in the electron momentum $\vec{k}$, a variety of SOFs are possible, depending on
the field direction. (Here, $\alpha$ is a system-dependent coefficient multiplied by the field strength,
and $\hat{E}$ denotes the field direction.) Application of the in-plane electric field, say $\hat{E}
\parallel +\hat{x}$, yields an SOF with the following form:
\begin{equation}\label{outofplane-Rashba}
\vec{\Omega}_{\textrm{SOF}}(\vec{k})=\alpha (\hat{x}\times \vec{k})=\alpha k_y \hat{z}.
\end{equation}
The SOF induced by the in-plane electric field shows unidirectional alignment perpendicular to the plane,
which has an identical form in the III-V quantum well grown along the [110] direction~\cite{Schliemann2017}.
This SOF may be referred to as the out-of-plane Rashba effect, in contrast with the conventional Rashba effect
that has in-plane SOF components~\cite{Bychkov1984} (see Fig. 1(a)). Combining the unidirectional
Rashba SOF with the quadratic kinetic energy yields the Hamiltonian $\mathcal{H}=\mathcal{H}_{\mathrm{kin}}
+\vec{\Omega}_{\textrm{SOF}} \cdot \vec{\sigma}=\frac{\hbar^2}{2m}(k_x^2+k_y^2)+\alpha k_y \sigma_z$, where
$\vec{\sigma}$ denotes the Pauli spin matrices. The in-plane electric field induces the two-dimensional
parabolic band to undergo an out-of-plane Rashba effect, leading to a spin texture identical to that seen
in the equivalent Rashba and Dresselhaus system, except for the spin quantization axis. Spin-up and spin-down
parabolic bands are shifted by a constant vector $\vec{Q}=\frac{2m\alpha}{\hbar^2}\hat{y}$, and a spin-degenerate
line node appears along the $k_y=0$ momenta parallel to the field direction.

When incorporated into the ferroelectricity, the in-plane electric field becomes accessible in a fully controllable
manner. The in-plane ferroelectricity naturally develops an in-plane electric field and produces a unidirectional
and maximally field-tunable Rashba SOF from a single inversion asymmetry. In general, the SOF induced by the
ferroelectric polarization can be written as~\cite{Disante2013,Kim2014}
\begin{displaymath}
\vec{\Omega}_{\textrm{SOF}}(\vec{k})=\alpha(\hat{P}\times \vec{k}),
\end{displaymath}
where $\hat{P}$, the polarization direction, replaces $\hat{E}$ in Eq.~\eqref{Rashba}.
Here, $\vec{\Omega}_{\textrm{SOF}}$ is completely locked on the ferroelectric polarization. Once again, if $\hat{P}$
is parallel to the plane, the unidirectional Rashba SOF described in Eq.~\eqref{outofplane-Rashba} emerges, but
with direct ferroelectric coupling. The ferroelectric-coupled out-of-plane Rashba SOF changes its sign by switching
its polarization. Moreover, a rich variety of controlled SOFs is anticipated, depending on the ferroelectric domain
structure. For example, in a target system composed of a square-based lattice, four 90$\,^{\circ}$-rotated configurations
of the in-plane ferroelectric moment are possible, and the out-of-plane Rashba SOF can vary accordingly (Fig. 1(b)).

As observed in the equivalent Rashba and Dresselhaus system, the out-of-plane Rashba band, in which the spin-up and
spin-down bands are shifted by a constant vector $\vec{Q}$, is known to host the long-lived PSH by recovering
the spin rotational SU(2) symmetry~\cite{Bernevig2006}. Injection of electron spin aligned with the plane induces
precession of the spin around the $z$-axis with a spatial periodicity of $l_{\mathrm{PSH}}\equiv \frac{2\pi}{|\vec{Q}|}
= \frac{\pi \hbar^2}{m \alpha}$ (Fig. 1(c)). In a practical application of the out-of-plane Rashba effect, the coherent
spin precession of the PSH can be exploited to fabricate a spin transistor with a size that is determined by
$l_{\mathrm{PSH}}$~\cite{Schliemann2003,Cartoixa2003,Kunihashi2012}. Because small spin transistors are required
for the development of high-density scalable spintronic devices~\cite{Sugahara2010}, the short pitch size of a
rapidly precessing PSH is highly desirable. A small $l_{\mathrm{PSH}}$ requires a large Rashba coefficient $\alpha$;
therefore, target two-dimensional ferroelectric materials would likely consist of heavy elements that possess large
atomic spin-orbit coupling.

\begin{figure}[b]
 \centering\includegraphics[width=8.5cm]{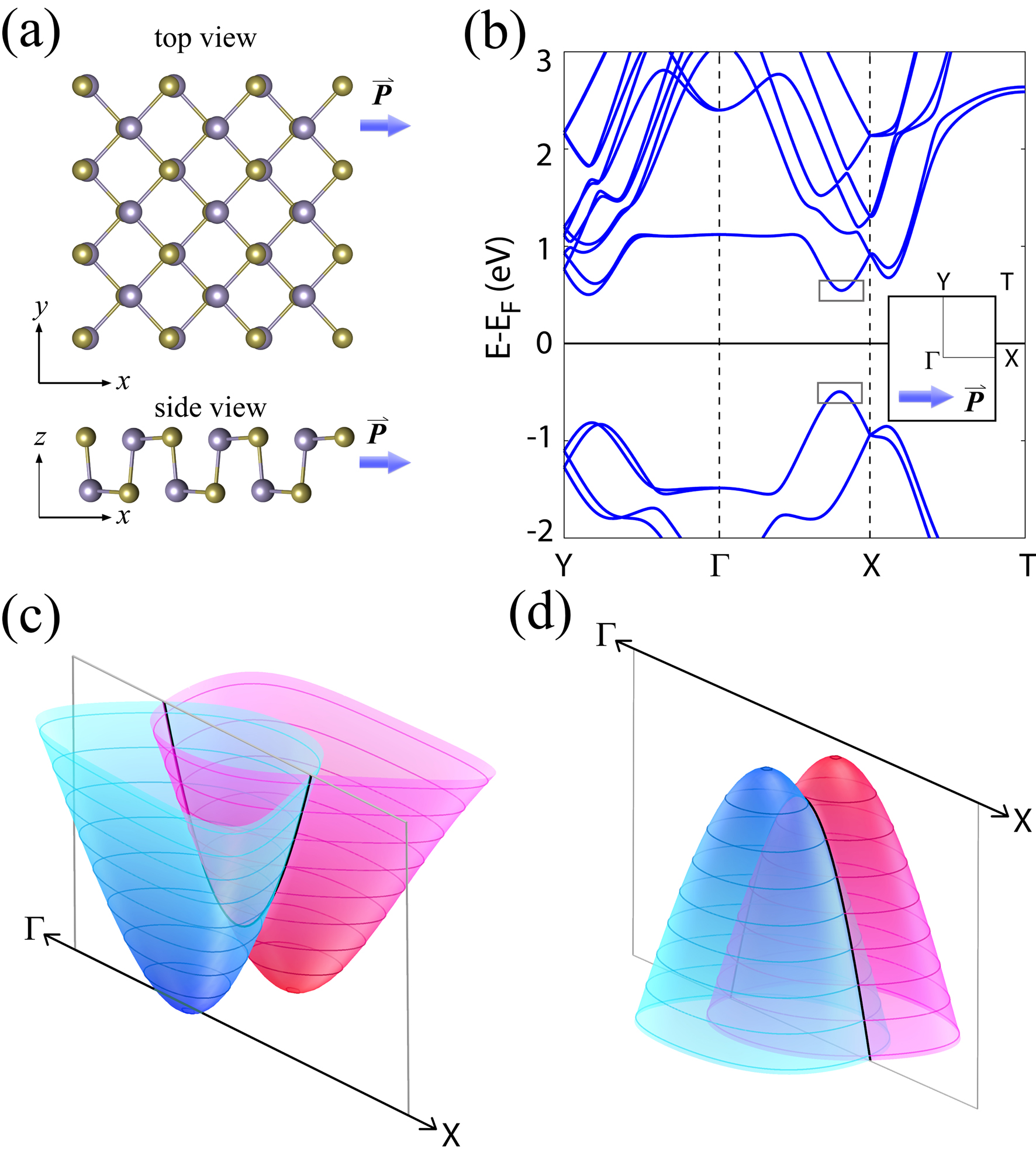}
 \caption{(a) Crystal and (b) electronic structures of the 1-unit-cell-thick SnTe film. Sn and Te atoms are colored
 blue and yellow, respectively. Ferroelectric-coupled out-of-plane Rashba bands emerged both at (c) the conduction band
 minimum (CBM) and (d) the valence band maximum (VBM). Spin-degenerate line nodes indicated by black solid lines in
 (c) and (d) correspond to the parabolic bands inside the grey box in (b). The interval between each contour line
 is 15 meV.}
 \label{fig:2}
\end{figure}

A (001)-oriented atomically thin SnTe film, recently synthesized using molecular beam epitaxy~\cite{Chang2016}, is
the most promising material potentially capable of satisfying the above criteria to produce a unidirectional,
field-tunable, and extremely large SOF driven by a single inversion asymmetry. To verify this idea, we have performed
density functional theory calculations, and found that the Heyd-Scuseria-Ernzerhof functional~\cite{Heyd2003} well
describes experimental ferroelectric displacements. For computational details and discussions on the lattice optimization,
see the Supplemental Material. As shown in Fig. 2(a), rock-salt based SnTe thin films display opposing displacements
of the tin and tellurium atoms within a plane, resulting in ferroelectric distortion parallel to the two-dimensional
plane. The overall electronic band structure of a one-unit-cell-thick SnTe film is shown in Fig. 2(b). The conduction
and valence band edges appear on the line $\Gamma$-X in the Brillouin zone, along which the ferroelectric polarization
is aligned. No spin splitting occurs near the band edges on the line $\Gamma$-X because this line corresponds to
the spin-degenerate line node seen in the out-of-plane Rashba SOF. By observing the low-energy electronic spectrum,
the out-of-plane Rashba band structures emerge, as shown in Figs. 2(c) and 2(d). Away from the spin-degenerate line
$\Gamma$-X, spin splitting appears such that the spin direction at each spin split-off band is normal to the SnTe
film. In other words, the unidirectional SOF enforces the out-of-plane spin texture. The contour plot of the spin
split-off bands reveals that the spin-up and spin-down bands close to each band edge overlap with a constant shift,
confirming the existence of a ferroelectric-coupled out-of-plane Rashba SOF in the SnTe thin films. (For electronic
structures of SnTe thin films with different thicknesses, see the Supplemental Material.)

In contrast to the conventional Rashba SOF, in which the time-reversal symmetry protects the spin degeneracy at
$\vec{k}=0$, extra symmetries are needed to realize an out-of-plane Rashba SOF in a crystalline solid. The lattice
symmetry should be compatible with the unidirectional spin texture and the spin-degenerate line node seen in the
out-of-plane Rashba band. Two mirror symmetries in a two-dimensional bulk crystal lacking inversion symmetry may be
a simple recipe for achieving an out-of-plane Rashba SOF. The system should be invariant under both in-plane mirror
and vertical mirror reflections (say $\mathcal{M}_{xy}$ and $\mathcal{M}_{zx}$, respectively). The unidirectional
SOF is protected by $\mathcal{M}_{xy}$ because only the out-of-plane SOF component, $\Omega^z_{\textrm{SOF}}$,
survives under the in-plane mirror reflection. In addition, $\mathcal{M}_{zx}$ regulates $\Omega^z_{\textrm{SOF}}$
as an odd function under the vertical mirror reflection; i.e., $\Omega^z_{\textrm{SOF}}(-k_y)=
-\Omega^z_{\textrm{SOF}}(k_y)$. (For more details, see the Supplemental Material.) Therefore, $\Omega^z_{\textrm{SOF}}$
is zero along the line $k_y=0$, and the spin-degenerate line node emerges along the intersection between the two
mirror planes. The in-plane ferroelectricity that coexists with two perpendicular mirror planes could potentially
provide a design principle guiding the implementation of a ferroelectric-coupled out-of-plane Rashba SOF. In SnTe
thin films, vertical mirror symmetry and an in-plane mirror reflection exist in combination with the half-unit-cell
translation to constitute a unidirectional SOF and a spin-degenerate line node~\cite{Appelbaum2016}. Symmetry
protection renders the SnTe thin films robust in their access to the out-of-plane Rashba SOF.

\begin{figure}
 \centering\includegraphics[width=8.5cm]{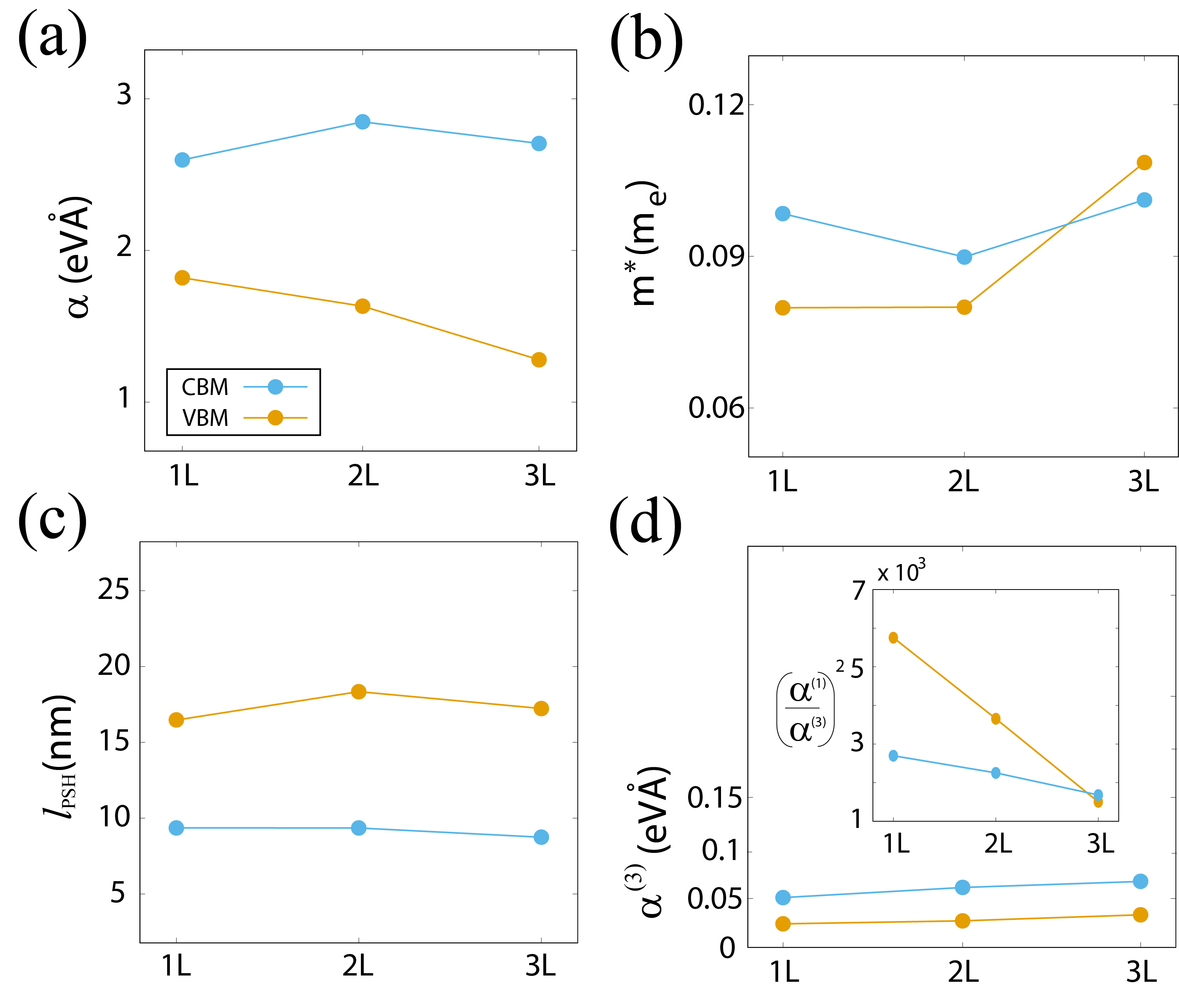}
 \caption{(a) The out-of-plane Rashba coefficient, (b) effective mass, (c) pitch size of PSH, and (d) $k$-cube Rashba
 coefficient at the CBM and VBM. 1L, 2L, and 3L denote one-, two-, and three-unit-cell-thick SnTe films, respectively.}
 \label{fig:3}
\end{figure}

The salient features of SnTe thin films for applications are illustrated in Fig. 3. Importantly, the giant Rashba
coefficients $\alpha$ of $1.28-2.85$ eV\AA ~are present in both the conduction and valence band edges (Fig. 3(a)).
These coefficients are two to three orders of magnitude larger than those in the III-V semiconductor quantum well
structures~\cite{Nitta1997,Koralek2009}. The giant value of $\alpha$ is attributed to large atomic spin-orbit coupling
and ferroelectric distortion in the SnTe thin films. Considering the effective mass shown in Fig. 3(b), the pitch size
of the PSH falls within the range $8.8-18.3$ nm (Fig. 3(c)). A huge reduction in $l_{\mathrm{PSH}}$ occurs compared
to the III-V semiconductor quantum wells that provided a PSH of a few micrometers in length~\cite{Koralek2009,Walser2012}.
Hence, the giant out-of-plane Rashba SOF improves the suitability of this material for use in nano-sized spin transistors
by handling the rapid but coherent spin precession of the PSH~\cite{Sugahara2010}.

The cubic out-of-plane Rashba term, $\alpha^{(3)}$, was estimated to investigate the long lifetime of the PSH in the
SnTe thin films (Fig. 3(d)). The higher-order momentum dependence of the SOF could be expressed as
\begin{eqnarray}
\Omega^z_{\textrm{SOF}}(\vec{k})&=&\alpha k_y + \alpha' k_x^2 k_y + \alpha'' k_y^3 \nonumber \\
&\simeq & \alpha^{(1)} k \sin{\theta} + \alpha^{(3)} k \sin{3\theta}, \nonumber
\end{eqnarray},
where $\alpha^{(1)}=\alpha +\frac{1}{4}(\alpha' + 3\alpha'') \langle k^2 \rangle$, $\alpha^{(3)}=\frac{1}{4}(\alpha' - \alpha'')\langle
k^2 \rangle$, and $\theta$ indicates the angle of momentum $\vec{k}$ with respect to the $x$-axis. By analogy to
the equivalent Rashba and Dresselhauss system, the main source of spin dephasing in the unidirectional SOF is the
$k$-cube term, which breaks the spin rotational SU(2) symmetry~\cite{Bernevig2006,Stanescu2007}:
$D_s \tau_{\mathrm{PSH}} \sim \left(\frac{\hbar^2}{m \alpha^{(3)}}\right)^2$, where $D_s$ is the spin-diffusion
constant, and $\tau_{\mathrm{PSH}}$ is the spin lifetime of the PSH mode~\cite{Walser2012}. Assuming a carrier
concentration $n=k_f^2/2\pi = 5\times 10^{11}$ cm$^{-2}$, we estimated $\alpha^{(3)} \simeq \frac{1}{4}(\alpha' - \alpha'') k_f^2$,
as plotted in Fig. 3(d). The inset of Fig. 3(d) shows $\left(\alpha^{(1)}/\alpha^{(3)} \right)^2$, which is
linked to the dimensionless quantity $\eta \equiv D_s \tau_{\mathrm{PSH}}|\vec{Q}|^2 \sim \left(\alpha^{(1)}/\alpha^{(3)} \right)^2$,
which describes the ratio of the spin relaxation times of the PSH and the ordinary spin diffusion~\cite{Koralek2009}.
The known $\eta$ values in the III-V semiconductors are of the order of $10^2$, and the overall values in the SnTe
thin films are $\sim 10^3$, indicating a slowly decaying PSH. By hosting a nanometer-sized and long-lived PSH,
the SnTe thin films may serve as an ideal platform for high-density spintronic devices that are immune to spin
dephasing.

\begin{figure}
 \centering\includegraphics[width=8.5cm]{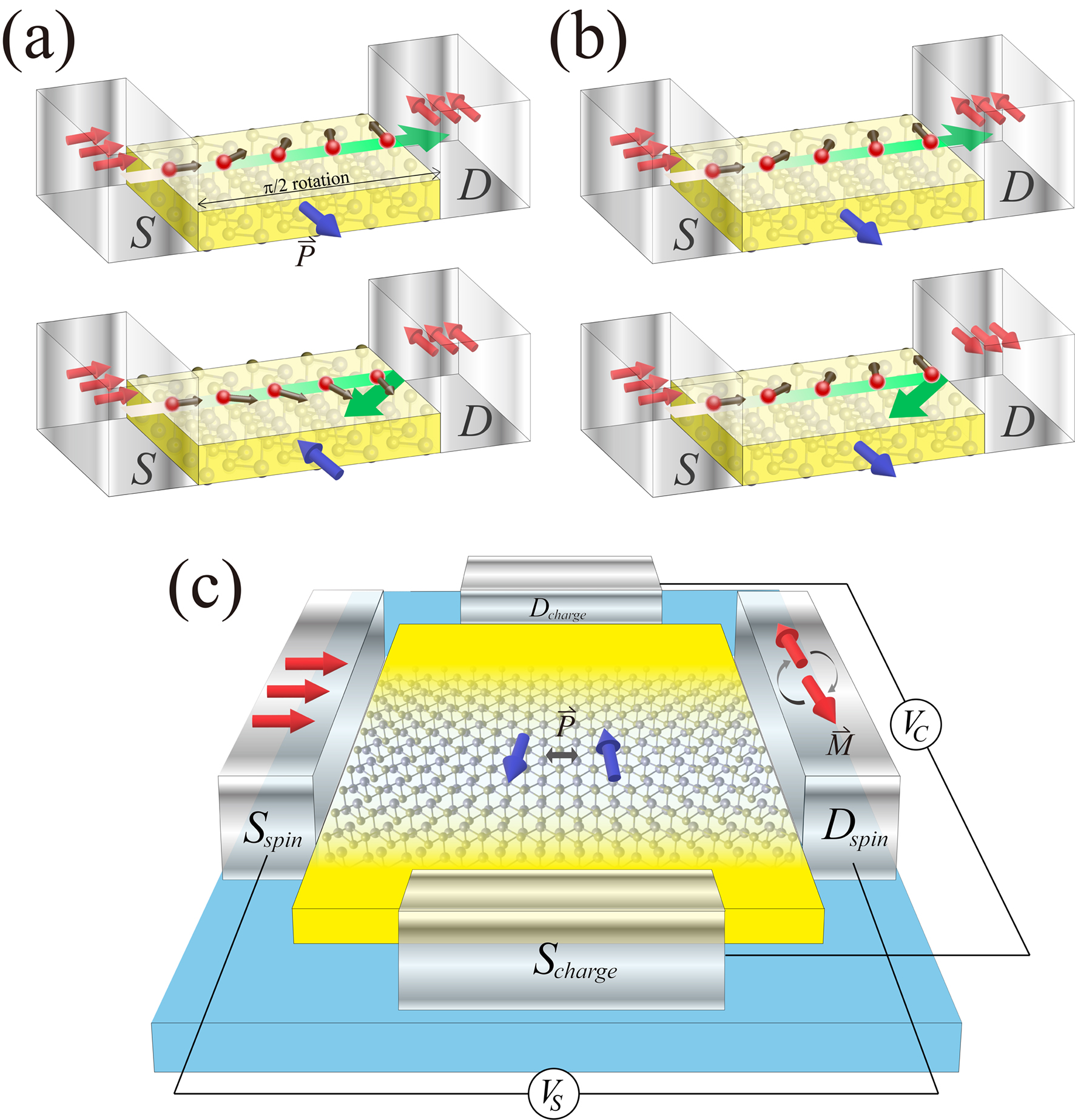}
 \caption{Two non-volatile on-off switching mechanisms are shown in the spin-valve-like structure, in which a SnTe
 thin film provides the spin transport channel: (a) electric and (b) magnetic on-off switching. (c) A multi-functional
 cross-shaped charge-spin transistor.}
 \label{fig:4}
\end{figure}

Considering the availability of non-volatile control over the ferroelectricity, SnTe thin films could enable novel
multi-functional spin transistors. As a simple sketch for a device application, a spin-valve-like experimental set-up
composed of ferromagnetic electrodes and a SnTe thin film as a transport channel could facilitate two switching modes
by employing a ferroelectric-coupled PSH. Ferromagnetic metal leads at both ends would have in-plane and orthogonal spin
orientations, and the ferroelectric polarization in the SnTe channel would align perpendicular to the transport direction.
The channel length would be set to $l_{\mathrm{PSH}}/4$ (modulo $l_{\mathrm{PSH}}/2$), corresponding to $\pi/2$ (modulo $\pi$)
spin precession of the PSH. Injection of the spin-polarized electron from the source electrode into the SnTe channel would
rotate the traversing spin around the $z$-axis by $\pi/2$ (modulo $\pi$) at the end of the channel~\cite{Kunihashi2016}.
Because the spin configurations of the ferromagnetic source and drain electrodes would be orthogonal, the electron spin
at the end of the SnTe channel would be either parallel or antiparallel to the spin orientation of the drain, leading to
an on- or off-state, respectively. The on-off switching could be controlled by flipping either the electric polarization
of the SnTe channel (Fig. 4(a)) or the magnetic polarization of the ferromagnetic electrodes (Fig. 4(b)). Combined with
the charge switching mechanism along the polarization direction~\cite{Chang2016}, we designed the multi-functional
cross-shaped charge-spin transistor sketched in Fig. 4(c). The intersectional charge and spin channels may be simultaneously
switched by controlling the electric polarization of the SnTe thin film, or the spin transport may be separately tuned
by adjusting the magnetic polarization of the ferromagnetic electrodes.

The unidirectional, ferroelectric-coupled, giant out-of-plane Rashba effect shown in SnTe thin films can provide
a new scheme for carrying and manipulating spin information, thereby broadening the scope of current spintronics
technologies. The nanometer lateral and atomically thin vertical scale over which the coherent spin is manipulated
could enable SnTe thin films to form heterostructures with other two-dimensional materials, such as van der Waals
layered compounds~\cite{Chang2016,Geim2013,Novoselov2016}, rendering them suitable for use in highly integrated
multi-functional devices. Additionally, given the superb thermoelectric response of bulk SnTe~\cite{Zhao2014,Zhao2016},
SnTe thin films could be a fascinating new option for thermo-spintronic applications~\cite{Bauer2012}, possibly
associating a drift in the PSH to an entropy-carrying degree of freedom under a temperature gradient.
\\

Financial support from the Basic Science Research Program of the National Research Foundation of Korea (NRF) under
Grant No. 2016R1D1A1B03933255 (H.J.) and 2017R1D1A1B03028004 (H.L.) is gratefully acknowledged.\\

\end{document}